\documentclass[10pt,showpacs,amsmath,amssymb,aps,floats,floatfix,nofootinbib]{revtex4}%
\usepackage[dvipdfm]{hyperref} % for latex & dvipdfm
\usepackage{epsfig}
\usepackage{dcolumn}
\usepackage{bm}
\usepackage{amsmath}
\usepackage{amsfonts}
\usepackage{amssymb}
\usepackage{graphicx}
\usepackage{color}
\usepackage{latexsym}
\usepackage{slashed} % slash mark
\def\fb{\mathrm{fb}} % fb
\def\ifb{\mathrm{fb}^{-1}} % fb^-1
\def\TeV{\mathrm{TeV}}     % TeV
\def\GeV{\mathrm{GeV}}     % GeV
\def\stop{{\tilde t}}     % stop
\def\tchi{{\tilde \chi}}  % neutralino or chargino
\def\stau{{\tilde \tau}}  % stau
\def\pT{p_\mathrm{T}} % p_T
\def\HT{H_\mathrm{T}} % H_T
\def\missET{\slashed E_\mathrm{T}} % missing E_T

\makeatother

\begin{document}
\title{Detecting light stop pairs in coannihilation scenarios at the LHC}
\author{Zhao-Huan Yu$^1$}
\author{Xiao-Jun Bi$^1$}
\author{Qi-Shu Yan$^2$}
\author{Peng-Fei Yin$^1$}
\affiliation{$^1$Laboratory of Particle Astrophysics,
Institute of High Energy Physics, Chinese Academy of Sciences,
Beijing 100049, China}
\affiliation{$^2$College of Physics Sciences,
University of Chinese Academy of Sciences,
Beijing 100049, China}

\begin{abstract}
In this work, we study the light stop pair signals at
the large hadron collider (LHC) in three coannihilation scenarios.
In order to yield the desired dark matter (DM) relic density, the
neutralino can coannihilate with stop, chargino and stau,
respectively. Signatures of the first scenario can be probed at
the LHC via the associated jet production processes $pp \to j +
{\tilde t} {\tilde t}^*$ by tagging an energetic mono-jet and a
large missing transverse energy. The signatures of the other
two scenarios can be searched via the pair production process $pp
\to {\tilde t} {\tilde t}^*$ by tagging energetic b-jets in the
final states and a large missing transverse energy. We find that
the LHC results at 7 TeV with 5 $\mathrm{fb}^{-1}$ of data can exclude
the stop mass up to 220, 380 and 220 GeV for these
three scenarios, respectively. While the 20 $\mathrm{fb}^{-1}$
dataset at 8 TeV is considered, the LHC can be expected to exclude the
stop mass up to 340, 430 and 370 GeV.
\end{abstract}

\pacs{12.60.Jv, 14.80.Ly}
\maketitle

\section{Introduction}

It is well known that DM plays a crucial role in the large scale
structure formation of the universe, however, its nature is still unclear.
Solving the nature of DM particle is a key problem in the cosmology and particle
physics.
There are many theoretical DM candidates proposed in literature \cite{Bertone:2004pz}, among which the lightest
supersymmetric particle (LSP) in the supersymmetry (SUSY) model with conserved
R-parity is a very attractive and widely studied candidate \cite{Jungman:1995df}.
%Many low scale SUSY models predict the lightest neutralino is the
Generally the lightest neutralino with mass
around 100 GeV up to TeV is the LSP in many low scale SUSY models.
It is possible to explore and test these models at the LHC.

The thermal relic density of neutralino which should be consistent with
the WMAP measurement \cite{Larson:2010gs} sets a strong constraint
on the SUSY parameter space.
If the neutralino is wino or higgsino dominated, the interactions between
neutralinos are strong enough to produce suitable DM relic
density. However, if the neutralino is bino-like, some additional
mechanisms are necessary to enhance the annihilation cross section
and avoid the over-production of neutralino.
For instance, the neutralinos can annihilate via a resonance
with fairly large cross section, which is the favored case
in the minimal supergravity model (mSUGRA) \cite{Fowlie:2012im}.
Except the resonance enhancement, coannihilation is another possible mechanism
to enhance neutralino annihilation rate and
produce the suitable DM relic density \cite{Griest:1990kh,Edsjo:1997bg}.
Typically, efficient coannihilation requires the LSP neutralino and the
next-to-lightest supersymmetric particle (NLSP) are nearly degenerate in mass.
In mSUGRA there are some typical parameter regions with significant coannihilation effects.
One is the low $m_0$ region where the stau can
be quasi-degenerate with the neutralino in mass
and coannihilates with the neutralino \cite{Ellis:1998kh}.
Another region is the so-called focus point region, where
the neutralino is a bino-higgsino mixture. In this region the chargino can be
light enough and induce a significant coannihilation effect~\cite{Edsjo:1997bg}.

It is interesting to notice that the colored superpartners,
like gluino \cite{Profumo:2004wk}, stop \cite{Boehm:1999bj} and sbottom \cite{Profumo:2003ema},
can also be the coannihilating partners of the LSP in the minimal supersymmetric
standard model (MSSM). In such cases, the strong interactions
between the NLSP particles can be more efficient to enhance the effective DM annihilation cross
section and to reduce the DM relic density. Moreover, the degenerate LSP-NLSP
spectra suggest the colored NLSP should also be light,
and such particles might have large production cross sections at the hadron colliders,
and can be within the reach of the LHC \cite{Ajaib:2010ne,Chen:2010kq,AdeelAjaib:2011ec,Bi:2011ha,Drees:2012dd}.

Recently, the LHC direct SUSY searches at $\sqrt{s}=7,8\,\TeV$ with
several $\ifb$ of data have set very strong bounds on the masses
of gluino and the squark of the first two generations \cite{:2012mfa,:2012rz}. However, if the colored NLSP is
almost degenerate with the LSP, the final states of the colored NLSP decay from direct pair
production are soft, and lead to a small reconstructed missing transverse
energy ($\missET$). Such signals are hardly triggered and selected by
the detectors. Therefore, considering the DM constraints, the LHC bounds
on the colored SUSY partner masses may be relaxed.
The light gluino, stop or sbottom can still
be consistent with the recent LHC results \cite{Ajaib:2010ne,Chen:2010kq,AdeelAjaib:2011ec,Bi:2011ha,Drees:2012dd}.

To overcome the triggering problem, an
additional energetic jet from initial state radiation is required. The new physics signals can be triggered by
this energetic mono-jet and the large $\missET$ \cite{Carena:2008mj,Fox:2011pm}. Although the cross section of this
mono-jet process is smaller than that of direct NLSP pair
production by a factor of ten or more if we require the
transverse momentum of mono-jet is larger than $100$ GeV, the
well-reconstructed large $\missET$ can efficiently
suppress the standard model (SM) backgrounds, especially the QCD
background. In literatures, such associated mono-jet production
channel has been demonstrated to be workable in various
coannihilation scenarios \cite{AdeelAjaib:2011ec,Drees:2012dd}.

In this work, we focus on the coannihilation scenarios with a
light stop quark. It is well-known that the lighter stop $\stop_1$
can be the lightest colored supersymmetric particle due to the
large top Yukawa coupling and large mass splitting terms in many SUSY models. The
light stop is also well-motivated by the ``naturalness'' argument
\cite{Kitano:2006gv}. Meanwhile, the electroweak baryogenesis
requires a light stop (say 100 GeV or so) to generate the first
order phase transition \cite{Carena:2008rt}. In order to
accommodate the data of the DM relic density, the
stop-neutralino coannihilation scenarios have also been proposed
in literatures. Recently, the LHC direct SUSY searches have set many
constraints on the light stop-neutralino mass spectra, but the
main results are only valid for the decay channel $\stop_1 \to t
\tilde{\chi}^0_1$ \cite{Aad:2012si,Aad:2012ar}. It is still
necessary to explore how the LHC can constrain the coannihilation
scenarios in the SUSY models.

In this work, we perform a more comprehensive study on the coannihilation scenarios with a light stop.
We consider the following three scenarios:
(1) $\stop_1$-$\tchi_1^0$ coannihilation scenario with $m_{\tilde{\chi}^0_1} \sim m_{\stop_1}$;
(2) $\tchi_1^\pm$-$\tchi_1^0$ coannihilation scenario with
$m_{\tilde{\chi}^0_1} \sim m_{\tilde{\chi}^\pm_1}<m_{\tilde{t}}$;
(3) $\stau_1$-$\tchi_1^0$ coannihilation scenario with $m_{\tilde{\chi}^0_1} \sim m_{\stau_1}< m_{\stop_1}$. For simplicity, the other supersymmetric particles are assumed to be much heavier here.

We use the associated mono-jet production channel to constrain the
parameter space for the first scenario. For the later two
scenarios \cite{Choudhury:2012tc,Baer:2012uy,Bi:2012jv}, we
consider the production channels $pp\to \stop_1\stop_1^* \to
b\bar{b}\tilde{\chi}^+_1 \tilde{\chi}^-_1$ and $pp\to
\stop_1\stop_1^* \to b\bar{b}\nu_\tau \bar{\nu}_\tau
\stau_1^+\stau_1^-$. The cross sections of the
electroweak supersymmetric particle direct productions, such as $pp \to \chi^0
\chi^{\pm}$ and $pp \to {\stau}{\stau}^*$, are much smaller than
those of the strong interaction. Therefore stop pair production $pp
\to {\tilde t} {\tilde t}^*$ can provide an advantage to probe
these two scenarios. Since the chargino/stau is nearly degenerate
with the LSP and the soft jets/taus from chargino/stau decay can
hardly be successfully reconstructed, we find that the latest
results from $\text{b-jets} + \missET$ searches can put
constraints on the parameter space. We would like to point out that the
constraints from the LHC on the stop signatures in these two scenarios are new and
have not been widely studied in literatures.

This paper is organized as follows.
In Section II, for the $\stop_1$-$\tchi_1^0$ coannihilation scenario,
we investigate the associated mono-jet production at the LHC and explore the constraints on the
parameter space by the latest $\text{monojet}+\missET$ searches.
We also explore the parameter region feasible with the $20\,\ifb$ dataset
at the collision energy $\sqrt{s} = 8\,\TeV$. In Section III,
we focus on two coannihilation scenarios with a light stop quark
where $\tchi_1^\pm$-$\tchi_1^0$ and $\stau_1$-$\tchi_1^0$ are almost degenerate, respectively.
We find that the searching channel of $\text{b-jets} + \missET$
can put constraints on the allowed parameter space.
We end this work in Section IV with discussions and conclusions.

\section{Stop-neutralino coannihilation scenario}

In this section we study the stop pair signature in the
$\stop_1$-$\tchi_1^0$ coannihilation scenario. The NLSP is assumed
to be the lighter stop $\stop_1$. Coannihilation with the LSP
$\tchi_1^0$ requires that $\stop_1$ is slightly heavier than
$\tchi_1^0$, saying $(m_{\stop_1}-m_{\tchi_1^0})/m_{\tchi_1^0}\lesssim 20\%$~\cite{Profumo:2004at}.
Then the decay modes $\stop_1 \to t \tchi_1^0$ and
$\stop_1 \to b W \tchi_1^0$ would be kinematically forbidden. The
loop-induced flavor changing neutral current decay mode $\stop_1
\to c \tchi_1^0$ becomes dominant, since the four-body decay modes
$\stop_1 \to f f' b \tchi_1^0$ are strongly suppressed by the
small phase space. Thus for the parameter region of $m_{\tchi_1^0}
+ m_c \leq m_{\stop_1} < m_{\tchi_1^0} + m_b + m_W$, we simply
assume the branching ratio of $\stop_1 \to c \tchi_1^0$ is 100\%.
Since the jets from stop decays are very soft and hardly
reconstructed, we consider the production channel of a stop pair
$\stop_1 \stop_1^*$ associated with at least one hard QCD jet.

Using the final states of $\mathrm{monojet} + \missET$ to search
for new physics, such as large extra dimension
and effective DM interaction operators, have been performed by
ATLAS~\cite{Aad:2012ky} and CMS~\cite{Chatrchyan:2012me}
collaborations at $\sqrt{s}=7\,\TeV$ with the integrated
luminosities of $4.7\,\ifb$ and $5.0\,\ifb$, respectively. The
crucial kinematic cuts used in these analyses are summarized in
Table~\ref{tab:monojet_cut}. Events in the $\mathrm{monojet} +
\missET$ channel are those contained large $\missET$ and an
energetic leading jet. Events with isolated leptons or more than
two jets with $\pT>30\,\GeV$ are rejected. The cuts
$\Delta\phi(\vec{j}_2,\vec{\slashed{E}}_\mathrm{T}) > 0.5$ and
$\Delta\phi(\vec{j}_1,\vec{j}_2) < 2.5$ are used to suppress QCD
multi-jet background events, where the large $\missET$ may come
from inefficient measurements of jets.
Table~\ref{tab:monojet_cut} also tabulates the corresponding
observed 95\%~CL upper limits on the beyond standard model (BSM) visible cross
section $\sigma_\mathrm{vis}^\mathrm{BSM} \equiv \sigma \cdot A
\cdot \epsilon$, which is defined as the production cross section
times acceptance and efficiency. In the following study, we apply
these latest limits to put bounds on the parameter space of the
$\stop_1$-$\tchi_1^0$ coannihilation scenario.

In our simulation, the parton-level events of SUSY processes $pp
\to \stop_1 \stop_1^* $, $pp \to \stop_1 \stop_1^* +
\mathrm{jets}$ and SM backgrounds are generated by
\texttt{MadGraph5}~\cite{Alwall:2011uj}.
\texttt{PYTHIA6}~\cite{Sjostrand:2006za} is used to perform parton shower,
particle decay and hadronization processes. Fast detector
simulation is carried out by \texttt{PGS4}~\cite{pgs}. The MLM
matching scheme with $\pT$-ordered showers implemented in
\texttt{MadGraph5} is adopted to overcome the parton-jet double
counting issue. For the stop production processes, the $Q_\mathrm{cut}$
parameter is chosen as 80 GeV which can yield the smooth jet
distributions. Jets are reconstructed by using the
anti-$k_\mathrm{T}$ clustering algorithm with a distance parameter
$R=0.4/0.5$ for ATLAS/CMS searches. The cross section of the stop
pair production including the NLO corrections is calculated by
\texttt{Prospino2}~\cite{Beenakker:1996ed}. For the top pair
production, the $K$-factor is calculated by \texttt{MCFM}~\cite{MCFM}.

\begin{table}[!htbp]
\begin{center}
\renewcommand{\arraystretch}{1.2}
\begin{tabular}{|c|c|c|c|}
\hline
 & ATLAS $7\,\TeV$, $4.7\,\ifb$ &
   CMS $7\,\TeV$, $5.0\,\ifb$ &
   LHC $8\TeV$, $20\,\ifb$ \\
\hline
Signal region  & SR1/SR2/SR3/SR4 & & \\
\hline
$\missET$ [$\GeV$] $>$ & 120/220/350/500 & 250/300/350/400 & 300 \\
\hline
$\pT^{j_1}$ [$\GeV$] $>$ & 120/220/350/500 ($|\eta|<2$) &
 110 ($|\eta|<2.4$) & 150 ($|\eta|<2.4$) \\
\hline
$\pT^{j_3}$ [$\GeV$] $<$ & 30 & 30 & 50 \\
\hline
 & $\Delta\phi(\vec{j}_2,\vec{\slashed{E}}_\mathrm{T}) > 0.5$ &
   \multicolumn{2}{c|}{$\Delta\phi(\vec{j}_1,\vec{j}_2) < 2.5$} \\
\hline
 & \multicolumn{3}{c|}{Lepton veto} \\
\hline
$\sigma_\mathrm{vis}^\mathrm{BSM}$ [$\fb$] $<$ &
1920/170/30/6.9 (95\% CL) & 120/73.6/31.6/19 (95\% CL) &
22.7/37.9 ($S/\sqrt{B} < 3/5$) \\
\hline
\end{tabular}
\end{center}
\caption{Crucial kinematic cuts in the ATLAS~\cite{Aad:2012ky}
and CMS~\cite{Chatrchyan:2012me} $\mathrm{monojet} + \missET$ analyses
as well as that in our $\mathrm{monojet} + \missET$ analysis
for $20\,\ifb$ at the LHC with $\sqrt{s}=8\,\TeV$ are tabulated.
The observed 95\%~CL upper limits on the BSM visible cross section
$\sigma_\mathrm{vis}^\mathrm{BSM}$
in the ATLAS and CMS analyses are provided,
as well as the expected upper limits on $\sigma_\mathrm{vis}^\mathrm{BSM}$
for $S/\sqrt{B} = 3$ and 5 in our analysis at $8\,\TeV$.
\label{tab:monojet_cut}}
\end{table}

The dominant SM backgrounds for this searching channel
are $Z (\to \nu \bar\nu) + \mathrm{jets}$ and
$W (\to \ell\nu) + \mathrm{jets}$. For the $W (\to \ell\nu) + \mathrm{jets}$ process,
charged leptons may be clustered into a jet or missed along the beam pile lines.
The irreducible background $Z (\to \nu \bar\nu) + \mathrm{jets}$ is most important.
To validate our MC results, we match the number of events of the
SM background in our simulation to those provided by the ATLAS and
CMS collaborations. We find that the corresponding rescaling
factors of our simulation are 1.05, 0.97, 0.92 and 0.86 for the
signal regions SR1, SR2, SR3 and SR4 in the ATLAS analysis, while
they are 0.98, 0.91, 0.96 and 0.97 for the signal regions with
$\missET > 250$, 300, 350 and 400$\,\GeV$ in the CMS analysis.
These rescaling factors are then applied to normalize the signal
events of $\stop_1 \stop_1^* + \mathrm{jets}$ in our simulation.

\begin{figure}[!htbp]
\begin{center}
\includegraphics[width=0.47\textwidth]{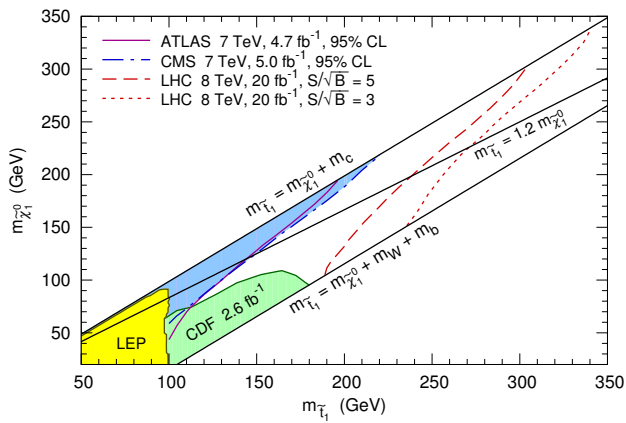}
\caption{The 95\% CL exclusion limits by ATLAS and CMS
$\mathrm{monojet} + \missET$ analyses at $7\,\TeV$
and signal significances predicted at $8\,\TeV$
in the $m_{\stop_1}$-$m_{\tchi_1^0}$ parameter plane
of the $\stop_1$-$\tchi_1^0$ coannihilation scenario are provided.
In contrast, the yellow and green regions show the excluded regions
by LEP and CDF, respectively.
\label{stop_limit}}
\end{center}
\end{figure}

The observed 95\% CL exclusion limits on mass parameter space
in the $\stop_1$-$\tchi_1^0$ coannihilation scenario are presented in Fig.~\ref{stop_limit},
We show the regions excluded by the ATLAS and CMS analyses.
In contrast, we also show the yellow and green regions, which have been excluded
by LEP and CDF~\cite{CDF:stop}, respectively. Obviously, the LHC has demonstrated its unique
potential to probe the $\stop_1$-$\tchi_1^0$ coannihilation scenario.

As the mass splitting $m_{\stop_1} - m_{\tchi_1^0}$ becomes
smaller, the charm quarks from $\stop_1$ decays become less
energetic, and so do their resulting jets. The only energetic jet
must be produced from initial state radiation and is opposite to
the stop pair direction. Thus the large $\missET$ can be
successfully reconstructed and used to trigger the signals.
Obviously, for a fixed value of $m_{\stop_1}$, a smaller
$m_{\stop_1} - m_{\tchi_1^0}$ means more easily being detected by
the $\mathrm{monojet} + \missET$ analysis, as demonstrated in Fig.~\ref{stop_limit}.

The region between the two lines labelled by ``$m_{\stop_1} =
m_{\tchi_1^0} + m_c$'' and by ``$m_{\stop_1} = 1.2
m_{\tchi_1^0}$'' is considered as the so-called ``coannihilation
region'' and is excluded up to $m_{\stop_1} \simeq 150 - 220
\,\GeV$ by the LHC collaborations. The most stringent limit is put
by CMS and can reach up to $m_{\stop_1} \simeq 220\,\GeV$. The
results of ATLAS agree with those obtained by the CMS. Obviously,
with more dataset, like the dateset of 20$\,\ifb$ at $\sqrt{s}=8\,\TeV$,
more parameter region can be excluded, as shown in Fig.~\ref{stop_limit}.

\begin{figure}[!htbp]
\begin{center}
\includegraphics[width=0.47\textwidth]{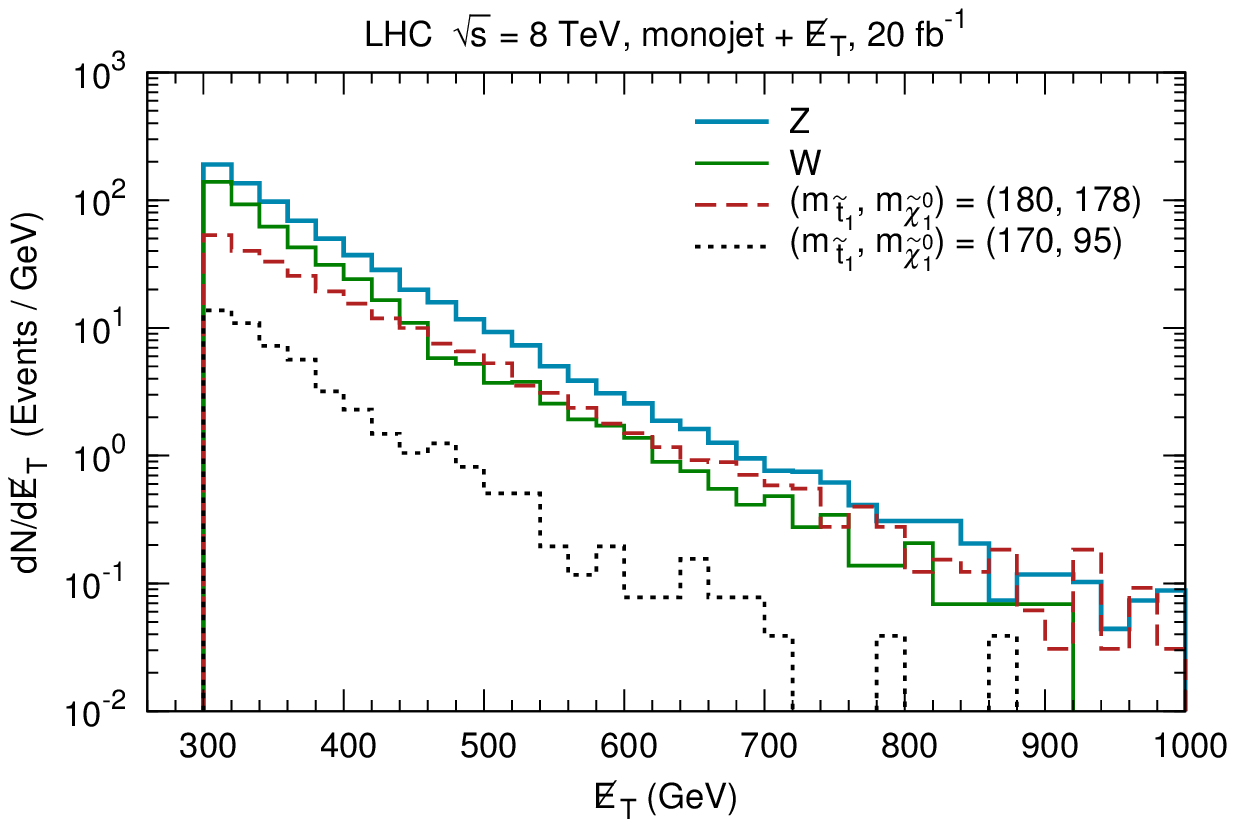}
\includegraphics[width=0.47\textwidth]{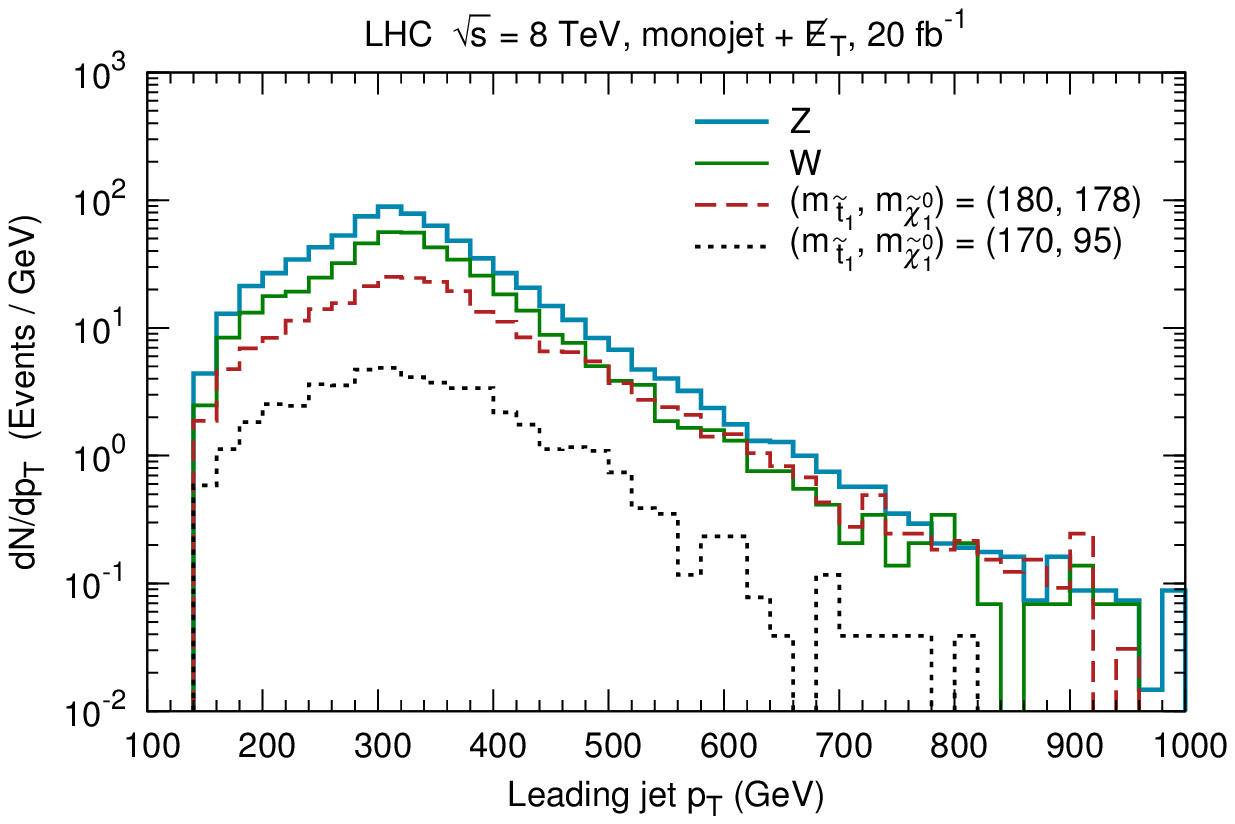}
\caption{The $\missET$ and the leading jet $\pT$ distributions
for the SM backgrounds and two benchmark signal points
in the $\stop_1$-$\tchi_1^0$ coannihilation scenario
at the LHC with $\sqrt{s}=8\,\TeV$ for $20\,\ifb$ are shown.
\label{stop_mEt_j1pt}}
\end{center}
\end{figure}

Now we look closely at our results for the LHC with a dataset of
$20\,\ifb$ at $\sqrt{s}=8\,\TeV$. After applying the cuts listed
in the last column of Table~\ref{tab:monojet_cut}, we arrive at a
total number of SM background events  $B=22944$ which includes
13939 $Z (\to \nu \bar\nu) + \mathrm{jets}$ and 9005 $W (\to
\ell\nu) + \mathrm{jets}$ events. With this number, we can have
the visible cross section of BSM $\sigma_\mathrm{vis}^\mathrm{BSM}
= 22.7\,\fb$ ($37.9\,\fb$) corresponding to the significance
$S/\sqrt{B} = 3$ (5).

\begin{figure}[!htbp]
\begin{center}
\includegraphics[width=0.47\textwidth]{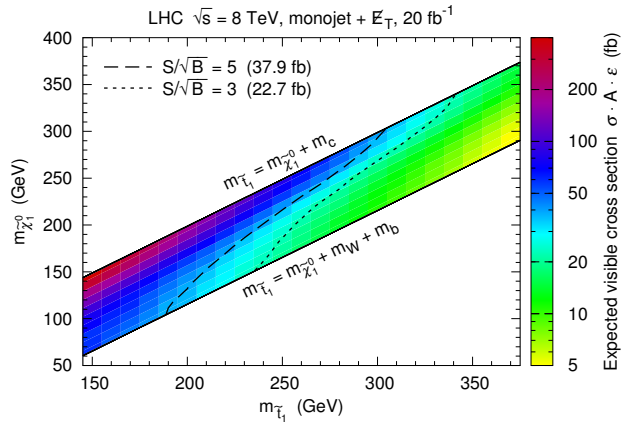}
\caption{Expected BSM visible cross section
in the $\stop_1$-$\tchi_1^0$ coannihilation scenario
at the LHC with $\sqrt{s} = 8\,\TeV$ for $20\,\ifb$ is shown.
The short (long) dash line corresponds to $S/\sqrt{B} = 3$ (5).
\label{stop_8TeV_color}}
\end{center}
\end{figure}

We also present the distributions of $\missET$ and the leading jet
$\pT$ in Fig.~\ref{stop_mEt_j1pt}, where the line shapes of two
main SM backgrounds and two benchmark points of SUSY model are
shown. In these two benchmark points,
$(m_{\stop_1},m_{\tchi_1^0})$ are deliberately chosen as $(180,
178)$ and $(170, 95)$ GeV to represent a nearly-degenerate and a
moderate splitting mass spectra, respectively. It is noticed that
all the lines in the leading jet $\pT$ distribution peak at $\sim
300\,\GeV$, which can be attributed to the fact that in most
events the $\missET$ approaches the cut threshold $300\,\GeV$ and
$\vec{\slashed{E}}_\mathrm{T}$ is mainly balanced by
the leading jet in the transverse plane.
We also find that the kinetic variable distributions of the
``nearly-degenerate'' benchmark point fall off more slowly than
those of the SM backgrounds. Nonetheless, increasing the cut
conditions of $\missET$ and the leading jet
$\pT$ can cause a sizable loss of signal with moderately suppressing the backgrounds. In order to
reduce the irreducible backgrounds, we have to go beyond these
kinematic cuts.

In Fig.~\ref{stop_8TeV_color}, we show the expected BSM visible
cross section in the $m_{\stop_1}$-$m_{\tchi_1^0}$ plane. The
short (long) dash line corresponds to $S/\sqrt{B} = 3$ (5) and
is also plotted in Fig.~\ref{stop_limit} for comparison. In
the ``coannihilation region'', the region with $S/\sqrt{B} > 3$ (5)
reaches $m_{\stop_1} \simeq 270 - 340\,\GeV$ ($240 - 300\,\GeV$).

\section{Chargino/stau-neutralino coannihilation scenario}

In this section, we study the stop pair signature $pp \to {\tilde
t} {\tilde t}^*$ in two coannihilation scenarios where $\tchi_1^0$
coannihilates with $\tchi_1^\pm$ and $\stau_1$, respectively. Here we would like to point
out that if these particles are mainly produced by stop decays the
LHC SUSY searches with tagged b-jets are suitable to test such
scenarios.

In the $\tchi_1^\pm$-$\tchi_1^0$ coannihilation scenario,
the NLSP chargino $\tchi_1^\pm$ is nearly degenerate with $\tchi_1^0$ in mass,
saying $(m_{\tchi_1^\pm}-m_{\tchi_1^0}) / m_{\tchi_1^0} \lesssim 20\%$~\cite{Profumo:2004at}.
We focus on the situation where the produced stop $\stop_1$ decays into $b \tchi_1^+$.
Because two-body decay channels of $\tchi_1^+$, such as $\tchi_1^+\to W^+ \tchi_1^0$ and $\tchi_1^+\to \nu \stau_1$,
would be kinematically closed, $\tchi_1^+$ dominantly decays into $\tchi_1^0$ and soft leptons or quarks.
In this case, the exact value of $m_{\tchi_1^\pm} - m_{\tchi_1^0}$ is of less importance
because the soft products from $\tchi_1^+$ decay can hardly be reconstructed by detectors.
Therefore, we fix $(m_{\tchi_1^\pm}-m_{\tchi_1^0}) / m_{\tchi_1^0}$ to be 10\%
as a typical case in the coannihilation scenario.
Furthermore, we assume
the branching ratio of $\stop_1 \to b \tchi_1^+$ is 100\%
in the region of $m_b + m_{\tchi_1^\pm} \leq m_{\stop_1} \leq m_{\tchi_1^0} + m_t$.

Although the decay products from $\tchi_1^+$ are soft, the $b$ quarks
from stop decays can be energetic enough to be tagged.
Meanwhile, the b-tagging technology is found to be powerful
to suppress the SM backgrounds.
Therefore the final state with $\text{b-jets} + \missET$
can be used to constrain this scenario.
The main SM background events are from production processes
of top pair, single top, $Z (\to \nu \bar\nu) + \text{heavy flavors}$
and $W (\to \ell\nu) + \text{heavy flavors}$.

\begin{table}[!htbp]
\begin{center}
\renewcommand{\arraystretch}{1.2}
\begin{tabular}{|c|c|c|c|}
\hline
 & ATLAS $7\,\TeV$, $2.05\,\ifb$ &
  \multicolumn{2}{c|}{ATLAS $7\,\TeV$, $4.7\,\ifb$} \\
\hline
Signal region  &  & SR2 & SR3a \\
\hline
$\missET$ [$\GeV$] $>$ & 130 & 200 & 150 \\
\hline
Leading jet & $\pT>130\,\GeV$, b-tagged & $\pT>60\,\GeV$, b-tagged
            & $\pT>130\,\GeV$ \\
\hline
2nd leading jet & $\pT>50\,\GeV$, b-tagged & $\pT>60\,\GeV$, b-tagged
                & $\pT>30\,\GeV$, b-tagged \\
\hline
3rd leading jet & $\pT<50\,\GeV$ & $\pT<50\,\GeV$
                & $\pT>30\,\GeV$, b-tagged \\
\hline
 & \multicolumn{2}{c|}{$m_\mathrm{CT}>100\,\GeV$} &
 $\Delta\phi(\vec{j}_1,\vec{\slashed{E}}_\mathrm{T}) > 2.5$ \\
\hline
 & \multicolumn{2}{c|}{$\Delta\phi(\vec{j}_{1,2},\vec{\slashed{E}}_\mathrm{T}) > 0.4$} &
 $\Delta\phi(\vec{j}_{2,3},\vec{\slashed{E}}_\mathrm{T}) > 0.4$ \\
\hline
 & \multicolumn{3}{c|}{$\missET / m_\mathrm{eff} > 0.25$, lepton veto} \\
\hline
$\sigma_\mathrm{vis}^\mathrm{BSM}$ [$\fb$] $<$ &
13.4 (95\% CL) & 2.29 (95\% CL) & 7.83 (95\% CL) \\
\hline
\end{tabular}
\end{center}
\caption{Kinematic cuts
in the ATLAS $\text{2 b-jets} + \missET$ analysis
for $2.05\,\ifb$~\cite{Aad:2011cw} and
in the updated analysis for $4.7\,\ifb$~\cite{ATLAS:2bjet_5ifb}
at the LHC with $\sqrt{s}=7\,\TeV$ are tabulated.
The observed 95\%~CL upper limits on $\sigma_\mathrm{vis}^\mathrm{BSM}$
are also given.
\label{tab:2bjet_cut}}
\end{table}

\begin{table}[!htbp]
\begin{center}
\renewcommand{\arraystretch}{1.2}
\begin{tabular}{|c|c|c|}
\hline
 & CMS $7\,\TeV$, $4.98\,\ifb$ &
  LHC $8\,\TeV$, $20\,\ifb$ \\
\hline
Signal region  & 1BL & $\geq$ 1BL\\
\hline
$\missET$ [$\GeV$] $>$ & 250 & 200 \\
\hline
 & $N_\mathrm{jet}(\pT > 50\,\GeV) \geq 3 $
 & $N_\mathrm{jet}(\pT > 60\,\GeV) \geq 3 $ \\
\hline
$\HT$ [$\GeV$] $>$ & 400 & 300 \\
\hline
 & $\Delta\hat{\phi}_{\min} > 4.0$
 & $\Delta\phi(\vec{j}_{1,2,3},\vec{\slashed{E}}_\mathrm{T}) > 0.4$ \\
\hline
 & \multicolumn{2}{c|}{At least one jet with $\pT>30\,\GeV$ tagged as b-jet} \\
\hline
 & \multicolumn{2}{c|}{Lepton veto} \\
\hline
 & - & $m_{jjj} \notin (130, 200)\,\GeV$ \\
\hline
$\sigma_\mathrm{vis}^\mathrm{BSM}$ [$\fb$] $<$ &
20.6 (95\% CL) & 8.4/14.0 ($S/\sqrt{B} < 3/5$) \\
\hline
\end{tabular}
\end{center}
\caption{Kinematic cuts
in the CMS $\text{b-jets} + \missET$ analysis~\cite{Chatrchyan:2012rg}
at $\sqrt{s}=7\,\TeV$,
and in our analysis for $\geq 1~\text{b-jets} + \missET$ final state
for $20\,\ifb$ at the LHC with $\sqrt{s}=8\,\TeV$ are tabluated.
The observed 95\%~CL upper limit $\sigma_\mathrm{vis}^\mathrm{BSM}$
in the CMS analysis is given,
so are the expected upper limits on $\sigma_\mathrm{vis}^\mathrm{BSM}$
for $S/\sqrt{B} = 3$ and 5 in our analysis at $8\,\TeV$.
\label{tab:1bjet_cut}}
\end{table}

The ATLAS collaboration has
reported the BSM searching in the $\text{2 b-jets} +
\missET$ channel at $\sqrt{s}=7\,\TeV$ with
$2.05\,\ifb$~\cite{Aad:2011cw} and
$4.7\,\ifb$~\cite{ATLAS:2bjet_5ifb} of data.
On the other hand,
the CMS collaboration has also released results in the $\text{b-jets} +
\missET$ channel at $\sqrt{s}=7\,\TeV$ with
$4.98\,\ifb$~\cite{Chatrchyan:2012rg} of data. The kinematic cuts
used in these analyses are summarized in Table~\ref{tab:2bjet_cut}
and \ref{tab:1bjet_cut}, where the crucial cuts in the signal
regions are tabluated. The observed 95\% CL upper limits on
$\sigma_\mathrm{vis}^\mathrm{BSM}$ are also given in the
tables.

Following the method proposed in Ref.~\cite{Fox:2011pm},
we estimate the observed 95\% CL upper limit
on the number of BSM events $N_\mathrm{BSM}$ by requiring
\begin{equation}
\chi^2 \equiv \frac{(N_\mathrm{obs}-N_\mathrm{SM}-N_\mathrm{BSM})^2}
{N_\mathrm{SM}+\sigma_\mathrm{SM}^2+N_\mathrm{BSM}} = 3.841,
\end{equation}
where $N_\mathrm{obs}$ is the number of observed events,
$N_\mathrm{SM}$ is the number of estimated SM background events,
and $\sigma_\mathrm{SM}$ is the uncertainty of $N_\mathrm{SM}$
including statistic and systematic uncertainties.
The 95\% CL upper limit on $\sigma_\mathrm{vis}^\mathrm{BSM}$
is obtained by dividing $N_\mathrm{BSM}$ by integrated luminosity.
For instances, the expected BSM visible cross sections
in the signal region SR2 of the ATLAS search
and in the signal region 1BL of the CMS search
are shown in Fig.~\ref{cha_SR2_1BL},
where the solid lines correspond to the observed 95\% CL exclusion limits.
The 95\% CL exclusion limits corresponding to
several signal regions in ATLAS and CMS searches at $7\,\TeV$
in the $m_{\stop_1}$-$m_{\tchi_1^\pm}$ plane
are also derived, as shown in Fig.~\ref{cha_limit}.

\begin{figure}[!htbp]
\begin{center}
\includegraphics[width=0.47\textwidth]{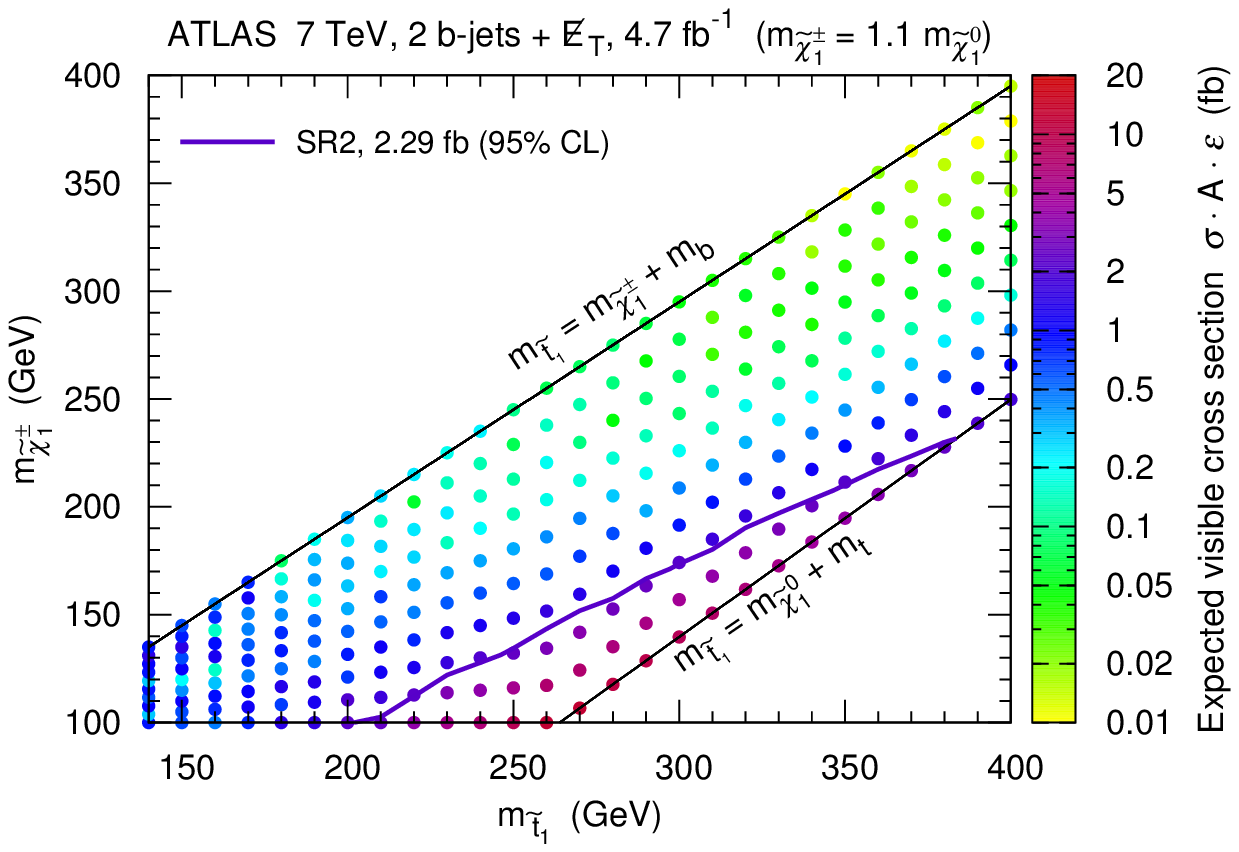}
\includegraphics[width=0.47\textwidth]{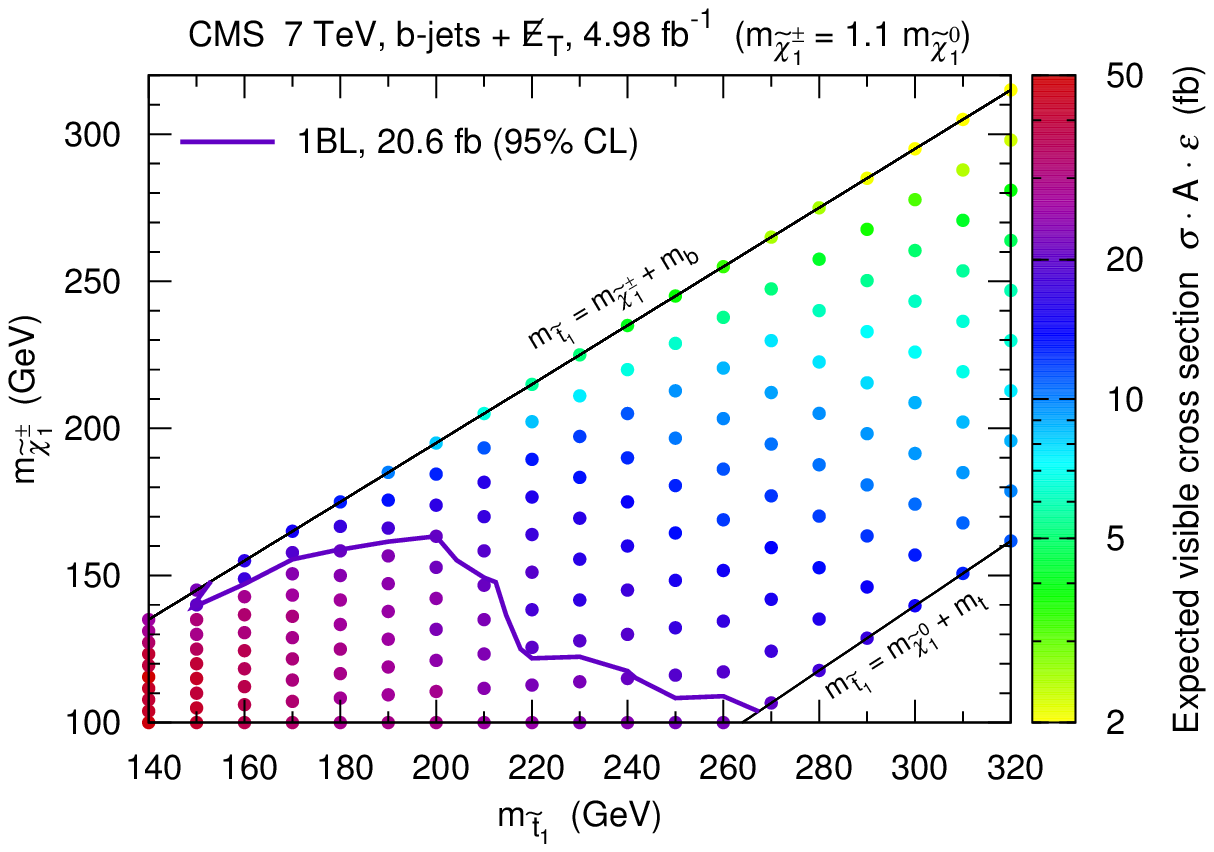}
\caption{Expected BSM visible cross sections
for the signal region SR2 of the ATLAS search (left frame)
and for the signal region 1BL of the CMS search (right frame)
in the $\tchi_1^\pm$-$\tchi_1^0$ coannihilation scenario
at $\sqrt{s} = 7\,\TeV$ are shown.
The solid lines correspond to the observed 95\% CL exclusion limits.
\label{cha_SR2_1BL}}
\end{center}
\end{figure}

Below we discuss the ATLAS analysis approaches in more details.
In the first and second approaches, the ATLAS searches focus on the
events with two b-jets and consider two types of kinematic
variables. 1) It requires
that there must be exactly two hard jets ($\pT>50 - 60\,\GeV$) and
both of them are b-tagged. To satisfy this requirement,
the mass splitting between $\stop_1$ and $\tchi_1^\pm$ should be
large enough. 2) In addition the contransverse mass
$m_\mathrm{CT} = \sqrt{(E_\mathrm{T}^{j_1}+E_\mathrm{T}^{j_2})^2
-(\mathbf{p}_\mathrm{T}^{j_1}-\mathbf{p}_\mathrm{T}^{j_2})^2}$
is required to be larger than 100 GeV. For stop pair events in the
$\tchi_1^\pm$-$\tchi_1^0$ coannihilation scenario, the
distribution of $m_\mathrm{CT}$ has an endpoint at
$m_\mathrm{CT}^{\max} =
(m_{\stop_1}^2-m_{\tchi_1^\pm}^2)/m_{\stop_1}$, which can be
achieved when the two b-jets are co-linear. Thus when the masses
of $\stop_1$ and $\tchi_1^\pm$ are close, stop pair events can
hardly satisfy the condition $m_\mathrm{CT} > 100\,\GeV$. On the
other hand, this condition also rejects quite many
top pair events of which the values of $m_\mathrm{CT}$
are often smaller than $100\,\GeV$.

It is found that these two types of cut conditions enable the first two analysis approaches
much easier to select out SUSY events with $m_{\stop_1}-m_{\tchi_1^\pm} \gtrsim 100\,\GeV$,
which can be read out in the left frame of Fig.~\ref{cha_SR2_1BL}.
For $m_{\stop_1} = m_{\tchi_1^0} + m_t$,
using the ATLAS searching result of the signal region SR2,
the 95\% CL exclusion limit can reach up to $m_{\stop_1} \simeq 380\,\GeV$.

\begin{figure}[!htbp]
\begin{center}
\includegraphics[width=0.47\textwidth]{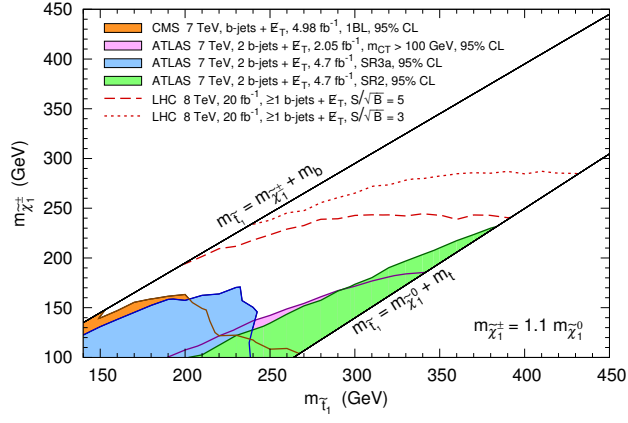}
\caption{95\% CL exclusion limits corresponding to
several signal regions in ATLAS and CMS searches at $7\,\TeV$
and signal significances predicted at $8\,\TeV$
in the $m_{\stop_1}$-$m_{\tchi_1^\pm}$ plane
of the $\tchi_1^\pm$-$\tchi_1^0$ coannihilation scenario.
\label{cha_limit}}
\end{center}
\end{figure}

In the third analysis approach, it requires a hard
leading jet which is not a b-jet, and the two b-jets can be soft. For
stop pair events, this approach tends to select the
events with an high $\pT$ jet from initial state radiation which
recoils against the stop pair system. Then this jet and the
missing transverse momentum are almost back-to-back, and the
condition $\Delta\phi(\vec{j}_1,\vec{\slashed{E}}_\mathrm{T}) >
2.5$ is useful to reduce SM backgrounds without losing many signal
events. As a result, the number of selected SUSY events mainly
depends on the production cross section and is not so sensitive to
$m_{\stop_1}-m_{\tchi_1^\pm}$. Therefore in Fig.~\ref{cha_limit},
the ATLAS searching result of the signal region SR3a excludes a
large region for $m_{\stop_1} \lesssim 240\,\GeV$.

\begin{figure}[!htbp]
\begin{center}
\includegraphics[width=0.47\textwidth]{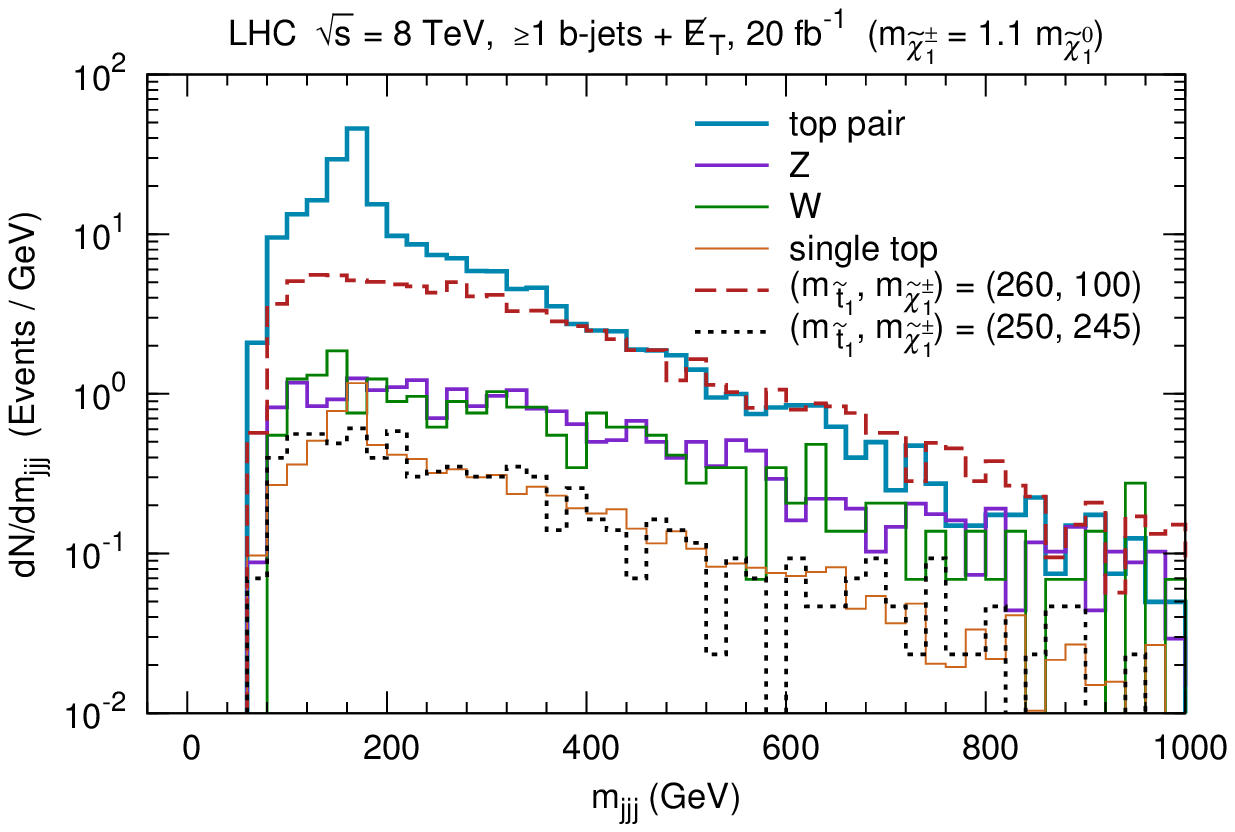}
\includegraphics[width=0.47\textwidth]{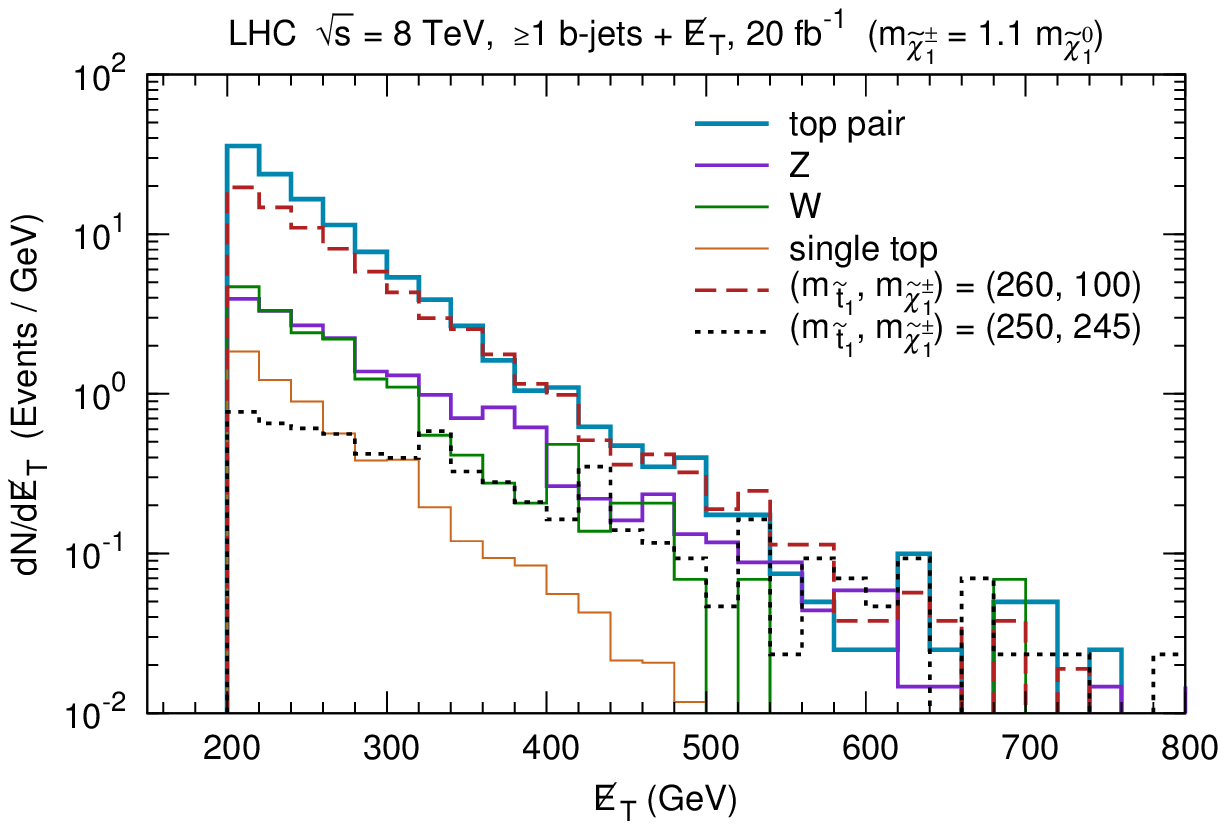}
\caption{The $m_{jjj}$ and $\missET$ distributions for the SM
backgrounds and two benchmark points in the
$\tchi_1^\pm$-$\tchi_1^0$ coannihilation scenario at the LHC with
$\sqrt{s}=8\,\TeV$ for $20\,\ifb$ are shown. The $m_{jjj}$
distributions are plotted before applying the condition $m_{jjj}
\notin (130, 200) \, \GeV$. \label{cha_mjjj_mEt} }
\end{center}
\end{figure}

The signal region 1BL in the CMS search focuses on
events with at least one b-jet.
It is required that there are at least three hard jets with $\pT>50\,\GeV$,
and the scalar sum of their $\pT$, $H_T$,
is demanded to be larger than $400\,\GeV$.
Events of stop pair production associating with initial state radiation jets
are easier to be selected.
The selection condition on b-jet is so loose that
just more than one b-jet with $\pT>30\,\GeV$ is required.
Consequently, for $m_{\stop_1} \lesssim 200\,\GeV$,
even the region where $m_{\tchi_1^\pm}$ closes to $m_{\stop_1}$
can be excluded, as shown in Figs.~\ref{cha_SR2_1BL} and \ref{cha_limit}.

\begin{figure}[!htbp]
\begin{center}
\includegraphics[width=0.47\textwidth]{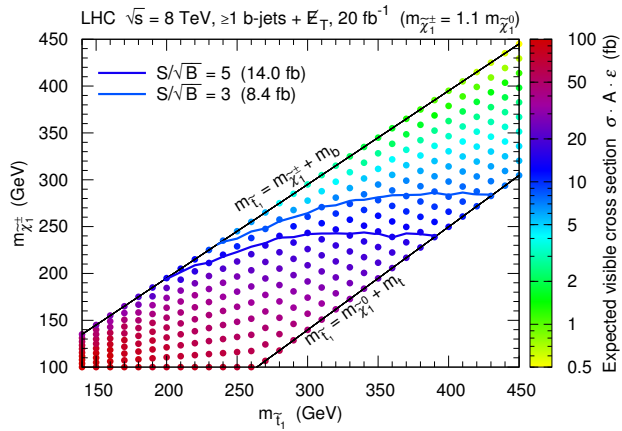}
\caption{Expected BSM visible cross section
in the $\tchi_1^\pm$-$\tchi_1^0$ coannihilation scenario
at the LHC with $\sqrt{s} = 8\,\TeV$ for $20\,\ifb$ are presented.
The solid lines correspond to $S/\sqrt{B} = 3$ and 5, respectively.
\label{cha_8TeV_color}}
\end{center}
\end{figure}

At the LHC with $\sqrt{s}=8\,\TeV$ and the luminosity of $20\,\ifb$,
we adopted the analysis approach listed in the second column of Table~\ref{tab:1bjet_cut}.
It is similar to the cut used in the signal region 1BL of the CMS search.
However, we choose lower cut conditions of $\missET$ and $H_T$
in order to select more signal events.

Furthermore, in order to suppress the $t{\bar t}$ background, we consider a
variable $m_{jjj}$. It has been used
in the ATLAS search~\cite{Aad:2012ar}, where it is applied to pick out hadronically decaying tops
instead of rejecting them. To construct $m_{jjj}$,
a pair of jets with invariant mass $m_{jj} > 60\,\GeV$ and smallest $\Delta R$
is picked out and reconstructed as a hadronically decaying $W$ boson.
A third jet which is closest to the reconstructed $W$ boson is also selected.
Then $m_{jjj}$ is the invariant mass of these three jets,
which may be the decay products of a hadronically decaying top.
In the events of top pair and single top production processes,
the distributions of $m_{jjj}$ nearly peak at
the mass of top quark $m_t = 173\,\GeV$.
This feature looks so clear in the left frame of Fig.~\ref{cha_mjjj_mEt},
where all the selection conditions in the kinematic cut are applied
except for the condition on $m_{jjj}$.
Then we can see that rejecting events with $m_{jjj} \in (130, 200)\,\GeV$
is useful to suppress top pair and single top backgrounds.
It can rejects 47\% (31\%) of top pair (single top) events,
while only rejects 20\% (21\%) of stop events
for $m_{\stop_1}=260\,(250)\,\GeV$ and $m_{\tchi_1^\pm}=100\,(245)\,\GeV$.
In the right frame of Fig.~\ref{cha_mjjj_mEt},
the $\missET$ distributions are also shown.

After applying all cuts, we arrive at a total number of SM background events $B=3132$
which includes 2269 top pair events,
390 $Z (\to \nu \bar\nu) + \text{hf}$ events,
353 $W (\to \ell\nu) + \text{hf}$ events
and 120 single top events.
Consequently, $S/\sqrt{B} = 3$ (5) corresponds to BSM visible cross section
$\sigma_\mathrm{vis}^\mathrm{BSM} = 8.4\,\fb$ ($14.0\,\fb$).
Fig.~\ref{cha_8TeV_color} shows
the expected $\sigma_\mathrm{vis}^\mathrm{BSM}$
in the $m_{\stop_1}$-$m_{\tchi_1^\pm}$ plane
of the $\tchi_1^\pm$-$\tchi_1^0$ coannihilation scenario.
The limits correspond to $S/\sqrt{B} = 3$ and 5
are plotted in both Figs.~\ref{cha_limit} and \ref{cha_8TeV_color}.

These limits are nearly horizontal
in the $m_{\stop_1}$-$m_{\tchi_1^\pm}$ plane
due to the following reason.
In general, when $m_{\stop_1}$ is fixed,
stop events with larger $m_{\stop_1}-m_{\tchi_1^\pm}$
are easier to induce hard jets and pass the kinematic cut.
When $m_{\stop_1}-m_{\tchi_1^\pm}$ is small, it needs
small $m_{\stop_1}$, which corresponds to large production cross section,
to yield more events to reach higher signal significance.
The region with $S/\sqrt{B} > 3\,(5)$ almost covers
all the space with $m_{\tchi_1^\pm} \lesssim 280\,(240)\,\GeV$
and reaches $m_{\stop_1} \simeq 430\,(390)\,\GeV$.

Finally, we study the $\stau_1$-$\tchi_1^0$ coannihilation
scenario. In many SUSY scenario stau can be light and less constrained
by direct search at the LHC. If stau is produced by decays of
charginos and heavier neutralinos, the isolated tau in the final
states is helpful to suppress the SM background. Especially, if
the final states of SUSY events contain tau and other charged
leptons or b-jets, the signatures can be well-reconstructed.
However, in the stau-neutralino coannihilation scenario, the NLSP
$\stau_1$ and the LSP $\tchi_1^0$ are nearly degenerate in mass,
the $\tau$ leptons from the stau decay $\stau_1 \to \tau
\tchi_1^0$ would be soft and can hardly be identified by
detectors. In this work, we fix the relation $m_{\stau_1} = 1.1
m_{\tchi_1^0}$. We focus on
the stop decay channel $\stop_1 \to b \stau_1^+ \nu_\tau$ and
assume that its branching ratio is 100\% for $m_b + m_{\stau_1}
\leq m_{\stop_1} \leq m_{\tchi_1^0} + m_t$.

\begin{figure}[!htbp]
\begin{center}
\includegraphics[width=0.47\textwidth]{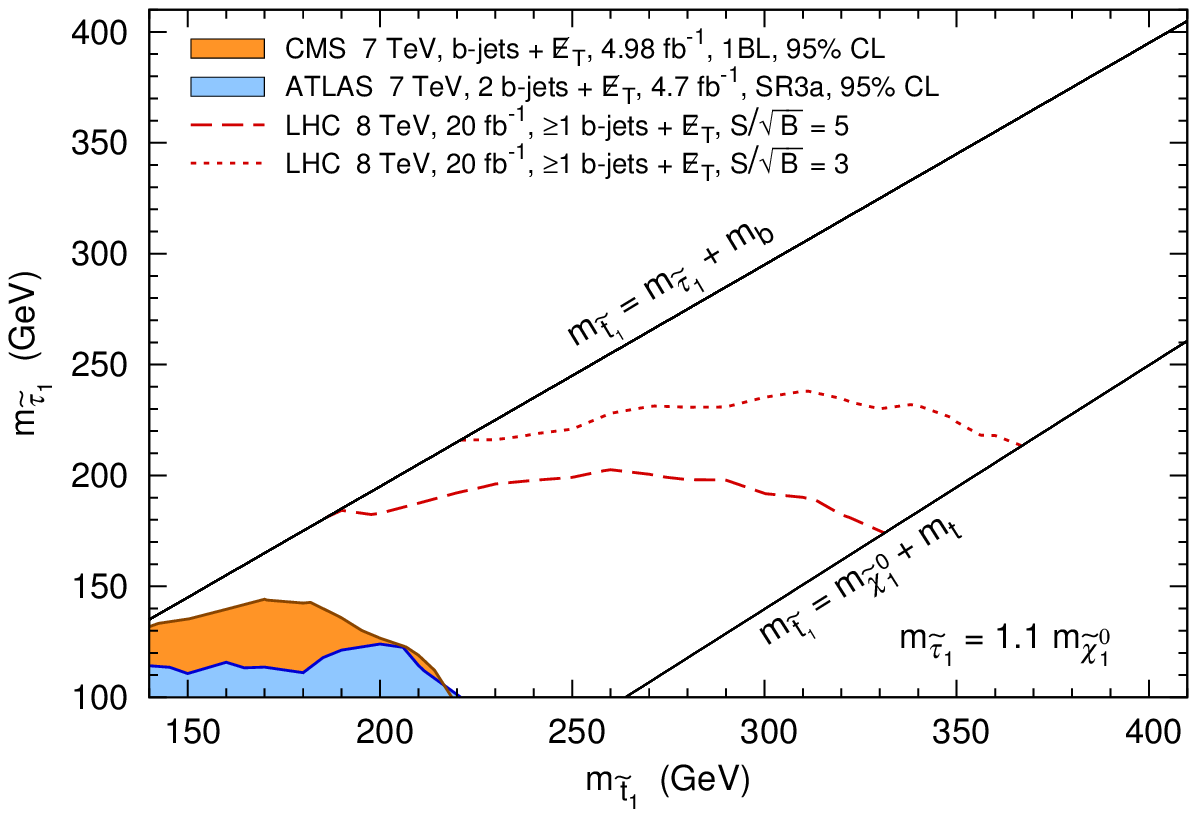}
\caption{The 95\% CL exclusion limits corresponding to
signal regions in ATLAS and CMS searches at $7\,\TeV$
and signal significances predicted at $8\,\TeV$
in the $m_{\stop_1}$-$m_{\stau_1}$ plane
of the $\stau_1$-$\tchi_1^0$ coannihilation scenario are provided.
\label{stau_limit}}
\end{center}
\end{figure}

The situation seems similar to that in the
$\tchi_1^\pm$-$\tchi_1^0$ coannihilation scenario. Due to the
3-body decay, however, the $b$ quark from stop decay tends to be
softer and it is not easy to be tagged. In addition, there are
four sources of missing transverse momentum, two neutrinos and two
neutralinos. The $\missET$ might be small when the momenta of
these four particles cancel out among themselves. Because of these
reasons, provided the same kinematic cuts, exclusion limits by
ATLAS and CMS searches at $7\,\TeV$ and signal significance
predicted at $8\,\TeV$ are all much weaker than those in the
$\tchi_1^\pm$-$\tchi_1^0$ coannihilation scenario, as demonstrated
in Fig.~\ref{stau_limit}. In the $m_{\stop_1}$-$m_{\stau_1}$
plane, the ATLAS and CMS searches just exclude a small region up
to $m_{\stop_1} \simeq 220\,\GeV$. At the LHC with
$\sqrt{s}=8\,\TeV$ for $20\,\ifb$, using the same kinematic cut we
adopted before, the region with $S/\sqrt{B} > 3\,(5)$ reaches
$m_{\stau_1} \simeq 240\,(200)\,\GeV$ and
$m_{\stop_1} \simeq 370\,(330)\,\GeV$.

\section{Conclusions and discussions}

In this work, we investigate the impacts of coannihilation scenarios on the light stop searches at the LHC.
For the $\stop_1$-$\tchi_1^0$ coannihilation scenario, the best searching channel is $\text{monojet}+\missET$.
Using the latest LHC mono-jet results, the present excluded region is up to $m_{\stop_1} \simeq
150-220\,\GeV$ for the coannihilation condition
$(m_{\stop_1}-m_{\tchi_1^0}) / m_{\tchi_1^0} \lesssim 20\%$.
Comparing with previous studies \cite{AdeelAjaib:2011ec},
the present mass bound has been improved to higher value.
At the LHC with $\sqrt{s} = 8\,\TeV$ for $20\,\ifb$, the region with
$S/\sqrt{B} > 3$ is expected to reach up to $m_{\stop_1} \simeq
340\,\GeV$.

We would like to mention that the
$\tilde{\chi}^\pm_1$-$\tilde{\chi}^0_1$ and the $\tilde{\tau}_1$-$\tilde{\chi}^0_1$
coannihilation scenarios are very general DM coannihilation scenarios,
and their features are well studied in many SUSY models.
If neutralino has a large
higgino component with a small $\mu$ value the chargino should also have a large
higgino component and their masses can be nearly degenerate. This is the case
for the $\tilde{\chi}^\pm_1$-$\tilde{\chi}^0_1$ coannihilation. When neutralino
is bino dominant, $\tilde{\chi}^0_1$ usually has large mass gap with the
chargino. In this case it needs to coannihilate with the other NLSP. If there is
some kind of unification at high energy scales, squarks are generally heavy
since gauge couplings tend to lift their masses when running from
high energy scale to low scale.
In this case, the lightest scalar sparticle
at low energy usually is the lighter stau.
Therefore the  $\tilde{\tau}_1$-$\tilde{\chi}^0_1$ coannihilation is another
important and common case to give correct DM relic density.
It is very important to study these two coannihilation cases
and consider their implications for SUSY search at the LHC.

In the $\tchi_1^\pm$-$\tchi_1^0$ coannihilation scenario, we concentrate
on the process $pp\to \stop_1\stop_1^* \to
b\bar{b}\tilde{\chi}^+_1 \tilde{\chi}^-_1$, and put constraints on
the parameter space by using the LHC
$\text{b-jets}+\missET$ results. We find that the ATLAS and CMS searches
at $\sqrt{s} = 7\,\TeV$ can exclude a region up to $m_{\stop_1}
\simeq 380\,\GeV$. The region corresponding to $S/\sqrt{B}>3$
expected at $\sqrt{s} = 8\,\TeV$ for $20\,\ifb$ almost covers all
the space in the $m_{\stop_1}$-$m_{\tchi_1^\pm}$ plane for
$m_{\tchi_1^\pm} \lesssim 280\,\GeV$ and $m_{\stop_1} \lesssim
430\,\GeV$.

In the $\stau_1$-$\tchi_1^0$ coannihilation scenario,
we focus on the process $pp\to \stop_1\stop_1^*
\to b\bar{b}\nu_\tau \bar{\nu}_\tau \stau_1^+\stau_1^-$,
which is more difficult than the case in the $\tchi_1^\pm$-$\tchi_1^0$ coannihilation scenario,
due to the neutrinos from $\stop_1$ decay.
We find that only a small region up to $m_{\stop_1} \simeq 220\,\GeV$ is excluded
by $7\,\TeV$ LHC searches.
At $\sqrt{s} = 8\,\TeV$ for $20\,\ifb$,
the expected $S/\sqrt{B}=3$ limit reaches
$m_{\stau_1} \simeq 240\,\GeV$ and $m_{\stop_1} \simeq 370\,\GeV$.

\begin{acknowledgments}
This work is supported by the Natural Science Foundation of China under the grant NO. 11105157, NO. 11075169, No. 11175251 and NO. 11135009, the 973 project under grant No. 2010CB833000, and the Chinese Academy of Science under Grant No. KJCX2-EW-W01.
\end{acknowledgments}

\setcounter{equation}{0}
\renewcommand{\theequation}{\arabic{section}.\arabic{equation}}%

\end{document}